	\documentclass[two columns]{aa}
	\usepackage{graphicx}
 	\usepackage{natbib}
 	
 	\bibpunct{(}{)}{;}{a}{}{,}

\begin{document}
	\title{Asymmetries of the Stokes~$V$ profiles observed by \textit{HINODE} SOT/SP in the quiet Sun}
	\titlerunning{SOT/SP Stokes $V$ profiles in the quiet Sun}
		
	\author{B.~Viticchi\'{e}\inst{1,2} \and
		J.~S\'{a}nchez~Almeida\inst{3}}
	\authorrunning{Viticchi\'{e} \& S\'{a}nchez~Almeida}
	
	\institute{ESA/ESTEC RSSD, Keplerlaan 1, 2200 AG Noordwijk, The Netherlands
		\email{bartolomeo.viticchie@esa.int} \and
		Dipartimento di Fisica, Universit\`a degli Studi di Roma
		``Tor Vergata'', Via della Ricerca Scientifica 1, I-00133 Rome, 
		Italy \and Instituto de Astrof\'isica de Canarias E-38205 La Laguna,
		Tenerife, Spain \email{jos@iac.es}}

\begin{abstract}
{}
{A recent analysis of polarization measurements of \textit{HINODE} SOT/SP
in the quiet Sun pointed out very complex shapes of Stokes $V$
profiles. Here we present the first classification of the SOT/SP
circular polarization measurements with the aim of highlighting exhaustively the whole
variety of Stokes $V$ shapes emerging from the quiet Sun.}
{\texttt{k-means} is used to classify \textit{HINODE} SOT/SP Stokes $V$ profiles observed
in the quiet Sun network and internetwork (IN). We analyze
a $302\times162$~arcsec$^2$ field-of-view (FOV) that can be considered
a complete sample of quiet Sun measurements performed
at the disk center with $0.32$~arcsec angular resolution and $10^{-3}$ polarimetric sensitivity.	
This classification allows us to
divide the whole dataset into classes, with each class
represented by a cluster 
profile, i.e., the average of the profiles in the class.}
{The set of $35$ cluster profiles derived from the
analysis completely characterizes the SOT/SP quiet Sun measurements.
The separation between network and IN profile shapes is 
evident -- classes in the network are not present in the IN,
and vice versa. Asymmetric profiles are approximately $93$~\% 
of the total number of profiles.
Among these, about $34$~\% of the profiles are strongly asymmetric,
and they can be divided into three families: blue-lobe, red-lobe,
and $Q$-like profiles. 
The blue-lobe profiles tend to be associated with upflows (granules),
whereas the red-lobe and $Q$-like ones
appear in downflows (intergranular lanes).}
{These profiles need to be interpreted considering
model atmospheres different from a uniformly magnetized
Milne-Eddington (ME) atmosphere, i.e.,
characterized by gradients and/or
discontinuities in the magnetic field and velocity along the
line-of-sight (LOS).
We propose the use of cluster profiles as a standard archive        
to test inversion codes, and to check the validity and/or
completeness of synthetic profiles produced by MHD simulations.}
\end{abstract}

\maketitle
	
\section{Introduction}
\label{Intro}
	The Stokes parameters ($I$, $Q$, $U$, and $V$)
	provide the main tool to analyze the solar photosphere.
	They are indeed molded by the photospheric
	structure and for this reason can tell us about its
	thermal, magnetic and dynamical properties.

	When considering Stokes profiles measured
	in spectral lines that are sensible to magnetic fields
	via Zeeman effect, one could expect
	the circular polarization profiles ($V$) to be
	exactly antisymmetric with respect to a central
	wavelength of the transition, possibly
	shifted by plasma dynamics\footnote{For
	linear polarization measurements ($Q$, and $U$),
	profiles are expected to be symmetric
	\citep[see, e.g.,][]{LanDLan04}.}.
	Contrary to such an idealization, the
 	Stokes~$V$ profiles
	emerging from the solar photosphere present
	important asymmetries
	independently of whether the
	measurements
	are obtained with low spatial resolution
	\cite[e.g., in plage or network regions;
	for a review see][]{Sol93} or with high spatial
	resolution \citep[e.g.,][]{Sig99,SanALit00,DomC06,
	Vit11}, and both in network and in
	the interior of the network (IN).
	
	Stokes~$V$ profile asymmetries are well known to be
	generated by gradients along and across the LOS
	of both magnetic field and plasma dynamics. Moreover,
	to deal with strong asymmetries
	one cannot consider mild variations
	of the above cited quantities, instead fairly 
	large changes like 	
	discontinuities have to
	be invoked \citep[e.g.,][]{GroD00}.
	Discontinuities in model atmospheres can
	be produced in many different ways. They are found
	in canopy models \citep[e.g.][]{GroD88,Ste00}
	or in embedded flux tube models \citep[e.g.,][]{SolMon93}, 
	in which one or two magnetopauses are present, i.e., sharp transitions
	from a magnetized region to a field-free one. 
	Besides these
	models, one can consider the presence of many
	discontinuities along the LOS
	in solar magnetic fields as being
	intermittent over scales smaller than the photon mean free path
	at the photosphere ($\simeq100$~km); these are the 
	MIcro-Structured Magnetized Atmospheres
	\citep[MISMAs,][]{SanA96,SanA98}.
	Smooth variations of the atmospheric properties can also be invoked, 
	but large asymmetries require steep 
	gradients in these continuous variations
	\citep[e.g.,][]{SolPah88,SanA88}.
	The asymmetries are also partly caused by
	discontinuities across the LOS, 
	in other words, an important factor which can 
	introduce asymmetries 
	is the finite spatial resolution
	of observations.
	The way it works remains unclear because 
	rather than decreasing the asymmetries
	with improving angular resolution, the 
	asymmetries observed in the quiet Sun
	with $0.32$~arcsec seem to be
	more extreme than those obtained with $1$~arcsec
	\citep[c.f.][]{SanALit00,Vit11}.
	On the one hand, one could expect to
	observe asymmetries mainly caused by gradients along
	the LOS when increasing the spatial resolution
	of the observations. On the other hand, we still
	do not know over which scale the magnetic
	field changes in the solar photosphere, e.g.,
	in \citet{VogSch07} the magnetic field changes
	over approximately $10$~km, well below
	the spatial resolution of \textit{HINODE}.
	
	The arguments considered above must be taken into account
	when interpreting Stokes measurements through inversion codes
	\citep[e.g.,][]{SocN01}.
	Among these, the ones based on ME 
	hypotheses cannot deal with asymmetries,
	but are still very important since they
	allow us to perform fast analyses of 
	large datasets \citep[e.g.,][]{SkuLit87,OroSDelT07,
	OroS07}.
	On the other hand, several
	inversion codes that can fit and interpret
	asymmetric profiles are available. The MISMA
	inversion code \citep[based on MISMA hypotheses,][]{SanA97}
	has been successfully used to interpret strong
	asymmetries in polarization profiles
	emerging from the quiet Sun \citep{SanALit00,DomC06,Vit11}.
	Other suitable tools to interpret Stokes $V$
	asymmetries are the SIR code \citep{RuiCTorI92},
	or the SIRJUMP code \citep[a modified version of
	the SIRGAUSS code, see][]{BelR03}, which was already used
	for the interpretation of \textit{HINODE} data
	\citep{Lou09}.\\
	
	Since its launch in $2006$, the spectropolarimeter
	SOT/SP \citep[][]{Lit01,Tsu08} of the Japanese
	mission \textit{HINODE} \cite[][]{Kos07} allows the solar community
	to perform spectropolarimetry of the solar photosphere
	under extremely stable conditions.
	SOT/SP provides $0.32$~arcsec angular resolution
	measurements with $10^{-3}$ polarimetric sensitivity.
	Moreover, its wavelength sampling ($2.15$~pm~pixel$^{-1}$)
	is suitable to have a detailed sampling of
	the polarization profiles.
	Many inversion analyses of spectropolarimetry measurements
	performed by SOT/SP have been performed so far.
	In most cases the hypothesis of ME atmosphere
	has been adopted 
	\citep[e.g.,][]{OroS07b,AseR09,IshTsu09}.
	Recently, \citet{Vit11} provided the first
	interpretation of Stokes profile asymmetries
	revealed in the dataset analyzed by
	\citet{OroS07b} and \citet{AseR09}; the analysis
	was performed under MISMA hypotheses.	
	
	The aim of this paper is to further point out
	and quantify the complexity of the shapes of SOT/SP measurements
	already discussed in \citet{Vit11},
	but using a tool independent of any inversion. To do this, we 
	adopt a method that allows us to extract and characterize
	the typical shapes of the Stokes~$V$ profiles observed
	by SOT/SP in the quiet Sun. As will be emphasized
	in \S~\ref{Results}, a large part of these
	turned out to be strongly asymmetric.
	The main implication of this result is that
	refined inversion techniques should be
	considered in future analyses of SOT/SP data
	to extract further information on
	the structure of the solar photosphere. 
	Those techniques are too burdensome to be  
	viable for large datasets,
	but both the profile classes and families
	we present provide a 
	small, yet complete set that is
	suitable for detailed study. 

	The paper is organized as follows: we introduce
	the dataset selected for the
	analysis in \S~\ref{Data}; we describe
	the classification tool adopted to extract
	the typical shapes of the Stokes~$V$ profiles
	from the dataset in \S~\ref{Analysis}; the
	typical profile shapes are presented
	in \S~\ref{Results}; we consider
	possible physical scenarios that could
	explain these profiles in \S~\ref{Disc}, and reflect on
	further applications of the classes; finally,
	our conclusions are outlined in \S~\ref{Conc}.
		
\section{Dataset}
\label{Data}
	We analyzed Stokes $I$ and $V$ profiles
	of a $302\times162$~arcsec$^2$ portion of the 
	solar photosphere observed at disk center on 
	2007 March 10 between 11:37 and 14:34~UT.
	The spectropolarimetric measurements
	were taken by the SOT/SP instrument
	aboard \textit{HINODE} \citep[][]{Lit01,Tsu08} in the two \ion{Fe}{i} $630$~nm
	lines, with a wavelength sampling of $2.15$~pm~pixel$^{-1}$, and 
	a spatial sampling of $0.1476$~arcsec~pixel$^{-1}$ and $0.1585$~arcsec~pixel$^{-1}$
	along the east-west and south-north directions, respectively.
	The data reduction and calibration were performed
	using the \texttt{sp\_prep.pro} routine available in the
	\textit{SolarSoft} \citep{Ich08}.
	After correction for the gravitational
    redshift, we verified that the cores of the
    full-FOV average Stokes $I$ profiles coincide
    within the SOT/SP wavelength sampling with the cores of the two Fe I
    lines in the Neckel atlas \citep{Nec99}.
    Moreover, the average line cores vary across the
    scanning direction with a standard deviation
    of approximately $60$~m~s$^{-1}$, discarding the problem mentioned by
    \citet{SocN11}. This accuracy of the
    absolute wavelength scale suffices because we mostly
    use relative velocities, and the absolute scale
    is employed only to distinguish granules and
    intergranules (i.e., upflows and downflows).
	Using the polarization signals
	in continuum wavelengths, we estimated	
	a noise level of $\sigma_V\simeq 1.1\times10^{-3}~I_c$
	for Stokes~$V$ ($I_c$ stands for the average continuum intensity
	of the FOV).
	The dataset has been already analyzed by \citet{OroS07},
	\citet{Lit08}, \cite{AseR09}, and \citet{Vit11} to derive 
	the magnetic properties of IN and network 
	regions. Consistently with these works,
	here we focus on Stokes~$V$ profiles whose maximum amplitude is
	larger than $4.5\times\sigma_V$. A total of $535465$
	Stokes~$V$ profiles meet the selection criteria
	(i.e., $26$~\% of the total number of profiles in the dataset).
	Such a large number of profiles can be rightly considered
	a complete set of polarization signals representative of the
	different Stokes~$V$ profiles that can be observed in
	the quiet Sun at disk center with $0.32$~arcsec resolution
	and $10^{-3}$ polarimetric sensitivity.
\begin{figure*}
	\centering
	\includegraphics[width=18cm]{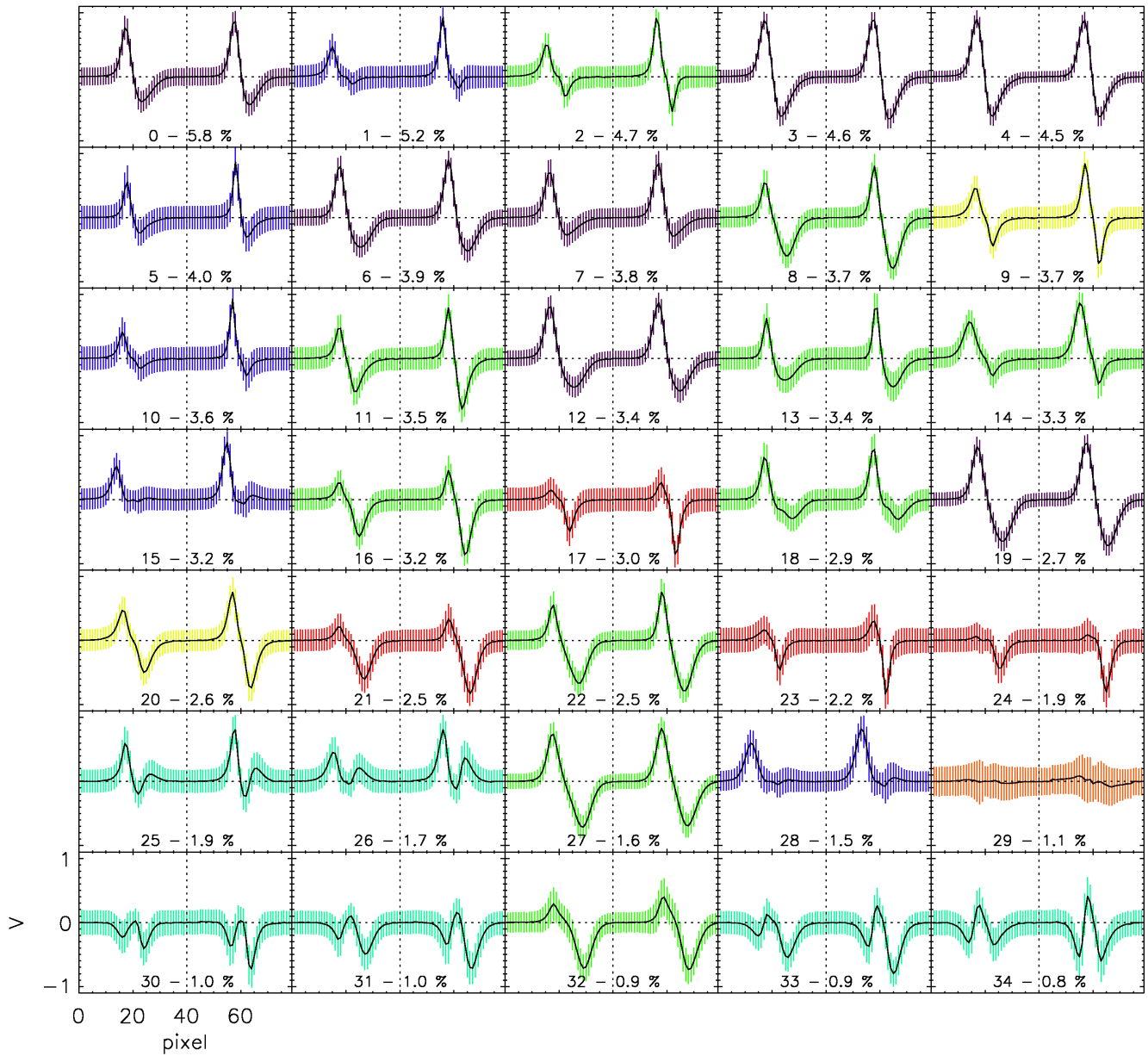}
	\caption{Classes of Stokes~$V$ profiles of \ion{Fe}{i}~630.15\,nm (left) and 
	\ion{Fe}{i}~630.25\,nm (right) in the quiet Sun according
	to a single \texttt{k-means} classification run.
	All the classes retrieved by the analysis are reported. 
	Each panel contains the average (solid lines) and the
	standard deviation (colored error bars) of all profiles in 
	the class. The dotted vertical line separates the two 
	spectral lines (it divides the $x$-axis in two windows of $40$ pixels, see \S~\ref{Dataarch}).
	Each plot is represented in the same $x-y$ range, i.e., the one of
	\texttt{class~30}. The number of the class is specified in each plot
	together with the percentages of profiles in the class. The
	colors group classes in families of alike profiles
	(\S~\ref{Results}): network (purple), blue-lobe (blue),
	red-lobe (red), $Q$-like (sky-blue), asymmetric (green), antisymmetric (yellow),
	and fake (orange).\label{fig1}}
\end{figure*}
\begin{figure*}
	\centering
	\includegraphics[width=15cm]{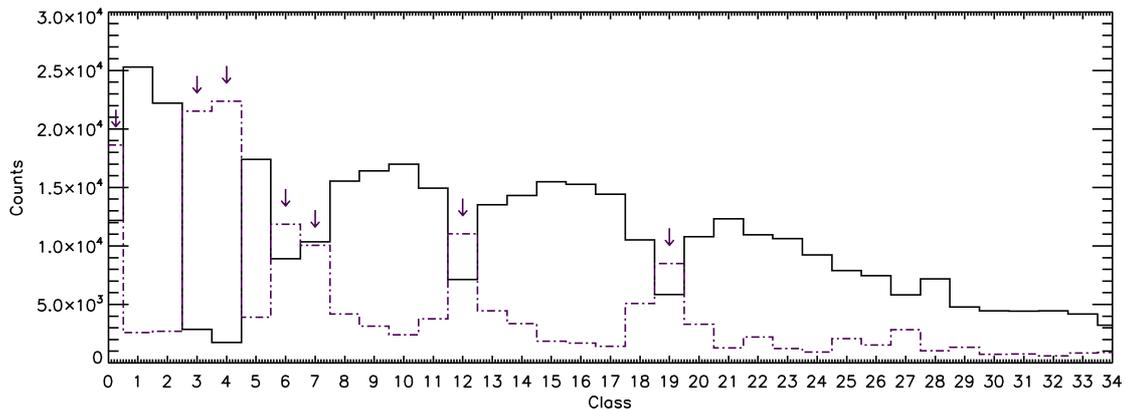}
	\caption{Histograms with the number of pixels in each class considering
		network regions (purple dot-dashed line) and IN regions (black solid line)
		separately. The arrows point out the network classes as derived from the
		statistics.\label{fig3}}
\end{figure*}
\begin{figure*}[tp!]
	\centering
	\includegraphics[width=22cm,angle=90]{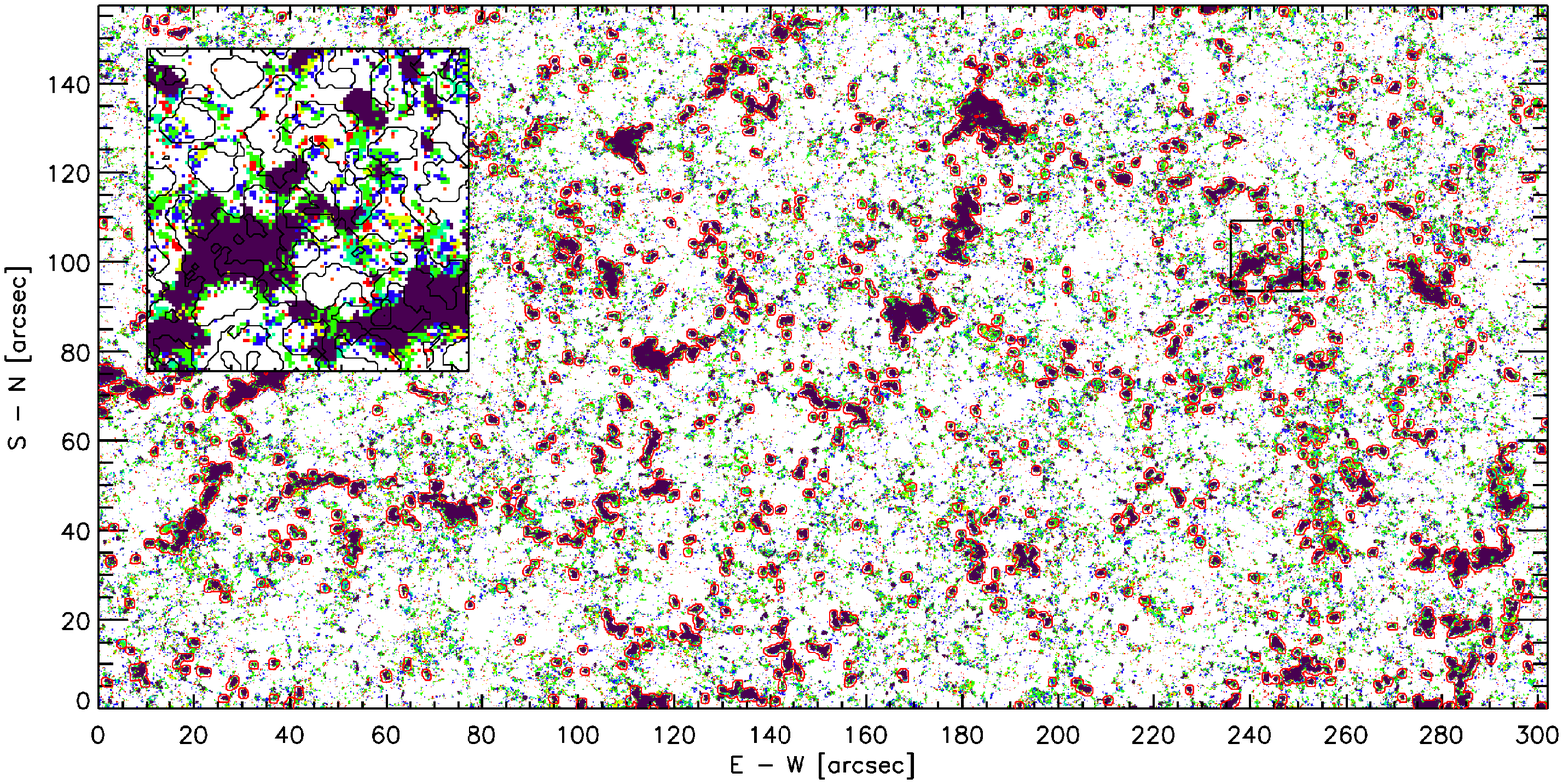}
	\caption{Spatial distribution of the profile families.
		Each color is
		representative of a single family according to the color code
		in Fig.~\ref{fig1}.
		White pixels are the ones with max$(\vert V \vert)<4.5\times\sigma_V$,
		which are excluded from the classification analysis. Red contours
		highlight the network regions in the FOV. The inset
		contains a $15\times 15$~Mm$^2$ detail of the FOV (black square) for a better
		visualization of the small family patches around a network
		region. The black contours highlight the
		regions with continuum intensity above the average continuum
		intensity of the FOV, i.e., the granules.\label{fig2}}
\end{figure*}

\section{Data analysis}
\label{Analysis}
	The analysis consists of three main parts.
	The most important one is the \texttt{k-means} classification
	of the profiles. This allows us to infer
	the typical profile shapes from the dataset described in \S~\ref{Data}.  
	Besides the classification, we derived the LOS
	velocity for each pixel in the 
	FOV adopting a Fourier transform method applied to both
	Stokes~$I$ lines separately \citep{TitTar75},
	and the center-of-gravity magnetogram signal from 
	Stokes $I$ and $V$ \citep[COG, ][]{ReeSem79}.
	
	The \texttt{k-means} classification
	algorithm is commonly used in data mining, machine learning,
	and artificial intelligence \citep[e.g., ][]{Eve95,Bis06}.	
	Indeed, the procedure we use is the same as the one used by \cite{SanA10} to
	analyze galaxy spectra from the Sloan Digital Survey Data
	Release 7. In this work the authors defined the procedure as
	{\em automatic} and {\em unsupervised}. 
	``Unsupervised'' implies that 
	the algorithm does not have to be trained,
	whereas ``automatic'' means that it is self-contained, 
	with minimal subjective influence. 
	In the standard formulation, \texttt{k-means} 
	begins by selecting at random from the full dataset 
	of $V$ spectra a number $k$ of template spectra. Each template spectrum 
	is assumed to be the center of a cluster, and each 
	spectrum of the data set is assigned to the 
	closest cluster center (i.e., that of minimum
	distance or, equivalently,
	closest in a least-squares sense). 
	Once all spectra in the dataset
	were classified, the cluster center is re-computed 
	as the average of the spectra in the cluster. This
	procedure is iterated with the new cluster centers, 
	and it finishes when no spectrum is re-classified 
	in two consecutive steps. 
	The number of clusters $k$ is arbitrarily chosen but,
	in practice, the results are insensitive to this selection
	because only a few clusters possess 
	a significant number of
	members, so that the rest can be discarded.
	On exit, the algorithm provides a number of 
	clusters, their corresponding cluster 
	centers, as well as the classification 
	of all the original spectra now 
	assigned to one of the clusters.
	The classes are ordered according to the total number of
	profiles they contain, i.e., assigning $0$ to the most abundant,
	$1$ to the second most abundant, and so on. 
	
	As a  major drawback, \texttt{k-means} yields different 
	clusters with each random initialization,
	therefore, one has to study this dependence
	by carrying out different random initializations
	and comparing their results.
	Partly trying to alleviate this problem,
	the code we employ uses an initialization 
	more refined than the plain random initialization.
	It consists of four steps:
	(Step 1) choose at random ten initial cluster 
	centers; (Step 2) run one iteration 
	of the standard \texttt{k-means}, 
	and select as initial cluster center 
	the cluster center with  the largest number 
	of elements; (Step 3) remove from the 
	set of profiles to be classified  those 
	belonging to the cluster center thus selected; 
	(Step 4) go to step 1 if profiles are still left;
	otherwise end. In order to show that the
	method is working properly, in \S~\ref{Results}
	we present the results obtained from a single
	classification run and then validate
	them by showing that they fully agree with
	the average results obtained when performing
	ten classification runs initialized with
	different random conditions.
	
	An important point to be mentioned
	is that the classes provided by 
	\texttt{k-means} do not form an orthogonal basis
	in the principal component analysis sense
	\citep[e.g.][]{Con95,Ree00}. The individual
	profiles are not unique linear superpositions of the cluster
	center profiles.
	Indeed, among the clusters found by the
	analysis, similar profiles can be easily recognized
	(Fig.~\ref{fig1}).
	However, it is important to specify that the aim
	of our classification is not to provide orthogonal
	classes, but to point out the typical shapes of the
	profiles observed by SOT/SP in the quiet Sun.
	The most important point to keep in mind is that
	all profiles associated to a  given
	class are similar to the cluster profile representative
	of the class and, therefore, the cluster profiles 
	are effectively representative of the shapes of the profiles
	observed by SOT/SP.
	Referring to the similarity between each cluster profile
	and the profiles associated to it, it is worth
	pointing out that profiles strongly deviating from the 
	cluster center are discarded. In more
	detail, when the cluster centers are calculated as the 
	average of the profiles in the class, those members
	with a distance larger than three times the standard deviation
	are rejected.

\subsection{Definition of the input dataset archive}
\label{Dataarch}
	Given a set of observations, the  actual data to be
	classified can be chosen in different ways (e.g., 
	range of wavelengths to be included, or whether 
	the profiles are corrected for the magnetic polarity).
	Here we explain the way our dataset was defined
	and why.

	The main idea is to analyze exclusively the shape
	of Stokes~$V$.
	For this reason we remove the global shift 
	of the profiles, renormalize each profile 
	to its maximum unsigned value, and sign the profiles 
	according to the main polarity in the resolution 
	element.
	
	Each profile is normalized to its maximum absolute
	amplitude as
	\begin{equation}
		\label{norm}
		V(\lambda)\longrightarrow V(\lambda)/\mbox{max}(\vert V(\lambda) \vert),
	\end{equation}
	where $\lambda$ spans over the whole wavelength
	range of SOT/SP (i.e., $0.24$~nm around $630.2$~nm.).

	Global wavelength shifts are removed separately for the 
	two spectral lines. We extract the Stokes~$V$ profiles of the 
	two \ion{Fe}{i} lines in two wavelength windows that are
	centered around the corresponding Stokes~$I$ line cores. 
	The two wavelength windows have
	the same amplitude, i.e. $86$~pm, so that both  
	lines {\em contribute} to the classification with
	the same number of wavelength pixels.
	In this procedure the number of wavelength pixels is reduced
	from $112$ to $80$ (i.e., two windows of $40$ wavelength pixels each).
	It is important to keep in mind that after this step it is not possible to
	consider a common wavelength range for the representation
	of Stokes~$V$ profiles of the two lines. For this reason,
	the abscissas of our Stokes~$V$ profiles are given in pixel units
	rather than wavelengths (see Fig.~\ref{fig1}).

	Finally, we individually sign the profiles according 
	to the polarity that dominates Stokes~$V$, so that if 
	$V$ were perfectly antisymmetric, the polarity-corrected
	Stokes~$V$ would have a positive 
	blue lobe and a negative red lobe. Since the signals are not 
	antisymmetric, we define {\em dominant polarity}
	as that yielding a $V$ with the sign of the
	largest average signal in the profile. In practice,
	we average Stokes~$V$ separately
	in the red-wing and the blue-wing 
	to decide which wing dominates. The 
	division between red and blue is set by the 
	Stokes~$I$ line core wavelengths. If the (unsigned)
	blue-wing signal is largest, then we divide
	Stokes~$V$ by its sign. If the (unsigned)
	red-wing signal is largest, then we divide
	Stokes~$V$ by minus its sign. The result is a $V$
	profile with the largest signal positive if 
	it happens in the  blue-wing, and negative if
	happens in the red-wing  
	(see the profiles in Fig.~\ref{fig1} for illustration).

\section{Results}
\label{Results}
\begin{figure}
	\centering
	\includegraphics[width=7cm]{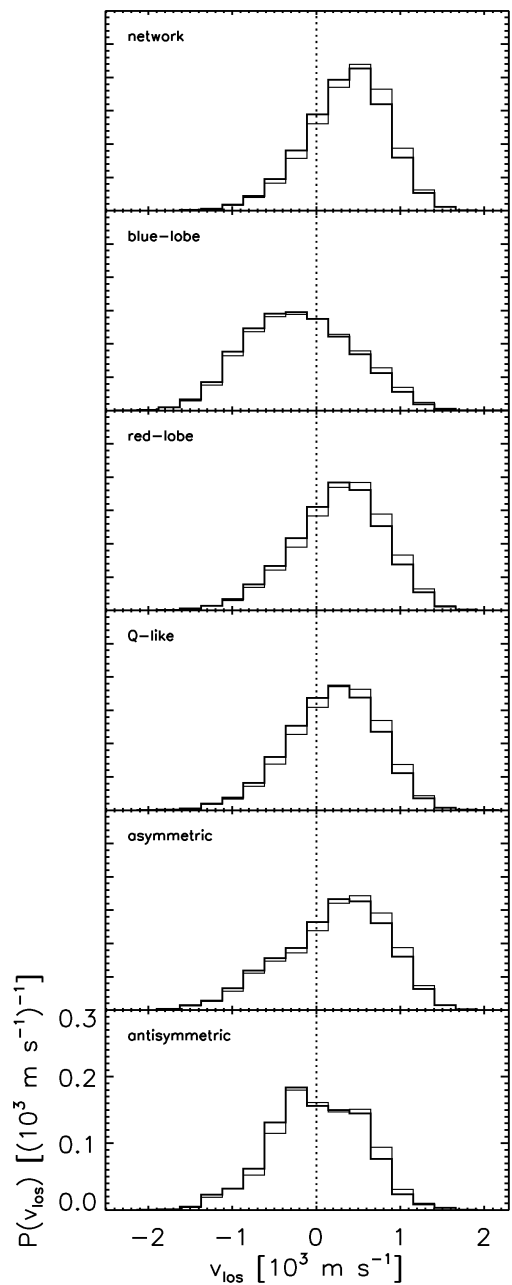}
	\caption{Statistics of the LOS velocities ($v_{los}$) of 
		the pixels associated with each family
		of Stokes $V$ shapes. Each plot contains the histograms of both \ion{Fe}{i}~630.15~nm 
		(thick lines) and  \ion{Fe}{i}~630.25~nm (thin lines).
		The histograms are normalized to the total number of profiles in the family.
		The vertical dotted lines separate
		the upflow regime (left) and the downflow regime (right).\label{fig4}}
\end{figure}
	The main result of the \texttt{k-means} classification
	is reported in Fig.~\ref{fig1}. It shows the cluster
	centers for all classes of profiles existing in the quiet Sun.
	After several trial classifications, we set the maximum number 
	of allowed classes to $50$, which exceeds 
	the typical number of classes
	retrieved by the procedure (i.e., around $35$, see Table~\ref{table}).
	In this way we are sure that the classification procedure is
	automatically setting the number of classes.
	The classification performed on the dataset of $535465$ profiles
	took about $20$~min on a standard laptop.	

	We number the classes according to the percentage
	of observed profiles belonging to the class, from the most
	common,  \texttt{class~0}, to the least common, \texttt{class~34}.
	It is important to notice that strongly asymmetric profiles are found
	throughout, e.g., the second most populated class
	shows important asymmetries. Classes of Stokes~$V$ with three lobes
	($Q$-like profiles) are also common, meaning that profiles like this are
	not isolated cases in quiet Sun measurements.
	
	As already reported in \S~\ref{Analysis}, similar profile
	shapes can be recognized among the cluster profiles
	in Fig.~\ref{fig1}. Indeed, attending to their basic properties
	the profiles can be grouped in six families: network, blue-lobe,
	red-lobe, $Q$-like, asymmetric, and antisymmetric \citep[partially
	following][]{Ste00}. With basic properties we refer to
	the major features of profiles, e.g., blue-lobe profiles have
	a very small red lobe so that, independently of the amount
	of polarization in the blue lobe, they are dominated by the blue lobe
	itself.
	
	The first family, i.e., the network one, can be defined
	by calculating the statistics for the \texttt{k-means} classes
	when network and IN regions are considered separately, i.e.,
	for the pixels included/not-included in the red contours of 
	Fig.~\ref{fig2}. The contours separate network and
	IN regions in the FOV. Network patches were identified
	with the algorithm used in \cite{Vit11}. We refer to this work for details,
	but it basically selects the patches with the largest polarization
	signals that all together cover some $10$~\% of the FOV
	\footnote{The network-IN separation is somewhat arbitrary.
	Another definition would make the network patches
	larger, including classes with larger asymmetries
	(see, e.g., the inset in Fig.~\ref{fig3}).}.
	The statistics is reported in Fig.~\ref{fig3},
	here the network classes are pointed out
	by arrows. More in detail, the network-IN division performed by the routine
	has an equivalent division in the \texttt{k-means} classes. Indeed,
	the classes that are abundant in the network (i.e., those
	pointed out by the arrows) are scarce in the IN,
	and vice-versa\footnote{One class is found to be shared by
	network and IN regions (i.e., \texttt{class~7}), we decided to consider
	it as a network class.}. These results imply that the network is
	associated with particular Stokes~$V$ shapes that are not found elsewhere.
	In other words, a classical network identification 
	and the \texttt{k-means} classification point out the same
	regions with two different methods: the first 
	one is mainly based on the analysis of the spatial coherence
	of strong Stokes~$V$ signals, whereas
	the second one does not take into account the amplitude
	of Stokes~$V$ (profiles are normalized before classification,
	see \S~\ref{Dataarch}), and refers just to profile shapes.
	The ``network family'' contains the following group of
	\texttt{k-means} classes: \texttt{classes~0}, \texttt{3},
	\texttt{4}, \texttt{6}, \texttt{7}, \texttt{12}, and \texttt{19}
	(the cluster profiles of these classes are
	reported in Fig.~\ref{fig1} with purple error bars).	
	These are representative of about $28.6$~\%
	of the total number of analyzed pixels.
	By examining the shapes of the clusters in the network family
	one can notice that these profiles present 
	the blue lobe larger than the red one, while
	the latter is more extended in wavelength (its area is larger than
	the area of the blue lobe).
	Such a shape is well known to be associated to network regions 
	\citep[e.g.,][]{Ste84,SanA96}.
	Moreover, the amplitude of the Stokes~$V$ profiles in the two lines is very similar
	in all network classes; this highlights the fact that a saturation due to
	kG fields is present, as expected in network regions
	\citep[see, e.g., ][]{SocNSanA02,Ste10}.
	As explained above, each profile emerging from \textit{HINODE} pixels
	with sufficient signal is associated to a class and,
	by grouping of classes through a certain criteria,
	to a family. In Fig.~\ref{fig2} we identify
	the position of a network profile with a
	purple pixel.
	The map shows how the network contours highlight
	large purple patches.
	
	On the other hand, clusters profiles in IN pixels present a huge variety
	of asymmetries. As an example, \texttt{class~1}, the most abundant class in the IN,
	is represented by a cluster profile dominated by the blue lobe. Many other classes are
	represented by profiles dominated by a single lobe, either the
	blue one or the red one (here always negative because of the condition
	imposed in \S~\ref{Dataarch}). The profiles in these classes
	represent approximately $34.4$~\% of the analyzed pixels.
	Among these, $17.5$~\% are the profiles
	dominated by the blue lobe, named the ``blue-lobe family'', i.e.,
	\texttt{class~1}, \texttt{5}, \texttt{10}, \texttt{15},
	and \texttt{28} (represented with blue error bars in Fig.~\ref{fig1}
	and blue pixels in Fig.~\ref{fig2})
	and $9.6$~\% by the red lobe, named
	the ``red-lobe family'', i.e., \texttt{17}, \texttt{21}, \texttt{23}, and \texttt{24}
	(represented with red error bars in Fig.~\ref{fig1}
	and red pixels in Fig.~\ref{fig2}).
	Moreover, a third family of extremely asymmetric profiles
	can be defined; \texttt{class~25}, \texttt{26}, \texttt{30}, \texttt{31},
	\texttt{33}, and \texttt{34} (represented with sky-blue error bars in Fig.~\ref{fig1}
	and sky-blue pixels in Fig.~\ref{fig2}) form the ``$Q$-like family'',
	which contains $7.3$~\% of the total number of analyzed pixels.
	
	The classes still not considered are IN profiles
	presenting shapes that are not clearly among the
	three families defined above. In spite of this,
	many of these profiles still present evident asymmetries;
	the ``asymmetric family'' is formed by the
	\texttt{class~2}, \texttt{8}, \texttt{11},
	\texttt{13}, \texttt{14}, \texttt{16}, \texttt{18}, \texttt{22},  
	\texttt{27}, and \texttt{32} which are $29.6$~\% 
	of the analyzed profiles (represented with green error bars in Fig.~\ref{fig1}
	and green pixels in Fig.~\ref{fig2}).
	Finally, only two classes show almost antisymmetric
	profiles; these represent only $6.3$~\% of the dataset
	(i.e., \texttt{class 9} and \texttt{20}) and are named the
	``antisymmetric family'' (represented with yellow error bars in Fig.~\ref{fig1}
	and yellow pixels in Fig.~\ref{fig2}). 
		
	Another interesting result comes from the study of the position
	of the families on the FOV. This can be done, for example, by studying
	the LOS velocity ($v_{los}$) for the pixels
	associated to each family.
	Pixels with $v_{los}<0$ ($v_{los}>0$) are
	placed in granules (intergranular lanes),
	i.e., upflow regions (downflow regions).
	Fig.~\ref{fig4} contains the result of such a study
	when calculating the statistics of $v_{los}$ (as
	derived by Stokes $I$, see \S~\ref{Analysis}) for
	the pixels included in the six families defined above.
	The main result derived from this study is that
	each family is related to a certain dynamics.	
	This is very revealing because the classification
	procedure does not know anything about $v_{los}$ because
	wavelength shifts were removed when extracting
	Stokes~$V$ profiles in the wavelength windows centered
	around Stokes $I$ cores (\S~\ref{Dataarch}).
	From this one learns that the plasma dynamics, as derived from Stokes $I$,
	i.e., the dynamics mostly associated to the non magnetized
	plasma in the pixel, is in some way related the shape of
	Stokes~$V$ profiles. One can elaborate a bit more on this result.
	The separation between the families
	with one lobe only (i.e., the blue-lobe and the red-lobe) is evident.
	On the one hand, profiles dominated by the blue lobe are mostly found in upflow regions,
	i.e., on top of the granules. On 
	the other hand, profiles dominated
	by the red lobe are mostly found in downflow regions, i.e.,
	intergranular lanes.
	The $Q$-like, network, and asymmetric families
	are found in downflow regions. Antisymmetric profiles
	are found in both upflow and downflow pixels with almost
	the same probability.
	Overall, as expected, most of the polarization signals reside 
	in downflows, i.e., $68.3$~\% of the analyzed pixels.
	
	Different families are also associated with different spatial extents. 
	When calculating the area of the largest patch
	\footnote{Here ``patch'' means a group of connected
	pixels belonging to the same family.} in the FOV 
	formed, exclusively, by profiles of the particular
	family one finds that network 
	patches, mainly associated with \texttt{class~3} 
	and \texttt{4}, can reach dimensions up to few
	tens arcsec$^2$. All other families have typical
	patch dimensions of few tenths of $1$~arcsec$^2$
	(see for a comparison the detail image in Fig.~\ref{fig2}
	in which contour highlighting the granular scale are reported).
	This means that out of network regions the profile
	shape of the analyzed profiles change over very
	small scales, smaller than the typical granular scale.
	
	The results presented in this section were derived from a single
	\texttt{k-means} classification run. As specified in \S~\ref{Analysis}
	it is important to study the dependence of the results
	on different random initializations. To do this, we derived
	the average abundances of the different families from ten
	classification runs; Table~\ref{table} allows one
	to compare the average results with those presented
	above. From this we can conclude that the results presented
	are sound and that the classification does not depend on the
	random initialization.
	
	The only class we did not consider so far in the presentation
	of the results is \texttt{class~29}, representative of $1.1$~\%
	of the analyzed profiles. This
	can be considered a fake class since it contains anomalous
	profiles. Here we define as anomalous, for example, those profiles
	on which the procedure adopted for the definition of the input dataset
	(\S~\ref{Dataarch}) did not work properly. A single fake
	class is found in all the classification runs; this is always
	representative of about $1$~\% of the analyzed profiles. This can
	be verified also in Table~\ref{table}, indeed by adding up the
	percentages from all families one gets $99$~\% of the
	analyzed profiles.
	
 	\begin{table}
	\caption{single run results vs. average results\label{table}}
	\centering 
	\begin{tabular}{lcc}
	\hline \hline
 	 & single run (Fig.~\ref{fig1}) & ten runs \\
 	\hline
		classes			&$35$		&$35\pm3$		\\
		network			&$28.6$~\% 	&$(28\pm3)$~\%	\\
		blue-lobe		&$17.6$~\% 	&$(15\pm2)$~\%	\\
		red-lobe		&$9.6$~\% 	&$(11\pm2)$~\%	\\
		$Q$-like		&$7.2$~\%	&$(8\pm2)$~\%	\\
		asymmetric		&$29.6$~\%	&$(30\pm5)$~\%	\\
		antisymmetric	&$6.3$~\%	&$(7\pm4)$~\%	\\
	\hline \hline
	\end{tabular}
 	\end{table}

\section{Discussion}
\label{Disc}
	An important point to emphasize
	is that the number of
	\texttt{k-means} classes representative of all
	the profiles observed by SOT/SP ($35\pm3$) is small
	compared to the total number of profiles
	in the analyzed dataset ($535465$). Moreover,
	when the major features of the classes are considered,
	the analyzed profiles can be organized in six
	families (i.e., network, blue-lobe,
	red-lobe, $Q$-like, asymmetric, and antisymmetric)
	so that	a few shape typologies can represent SOT/SP
	quiet Sun data. Nevertheless, it is important to note
	that almost all these shapes are strongly asymmetric.
	
	Most SOT/SP data have been so far interpreted through
	ME codes \citep[][]{OroS07,AseR09}, but
	the \texttt{k-means} classes in Fig.~\ref{fig1}
	show that such an hypothesis has to be replaced
	to analyze the whole variety of profile shapes observed
	by SOT/SP in the quiet Sun. 
	Except for the classes whose profile shapes can be reasonably
	considered to be quasi-antisymmetric (see \S~\ref{Results}),
	we can conclude that ME codes cannot reproduce 
	the observed profiles observed by SOT/SP.
	\citet{OroS10}
	recently showed that ME codes can be used to derive
	average properties of fully resolved magnetized
	atmospheres (technically speaking with magnetic
	filling factors equal to $1$) even when inverting
	extremely asymmetric profiles. This good news
	does not eliminate the need for more refined
	inversions. The asymmetries provide information
	on the unresolved structure and/or gradients along the
	LOS of the magnetic field that are washed out unless the
	line shapes are adequately interpreted. Moreover, the present
	angular resolutions do not grant treating all IN
	magnetic features as fully resolved.
	
	These considerations can
	be used in a constructive way when thinking
	about future strategies for the interpretation of
	spectropolarimetric data either from \textit{HINODE} or
	from future telescopes. One needs different hypotheses
	in order to interpret (and fit) the different polarization
	profiles emerging from the quiet Sun.
	As an example, let us consider two extreme cases
	from our \texttt{k-means} classification, e.g., \texttt{class~9}
	and \texttt{class~25}. The first one, quasi-antisymmetric,
	can be interpreted adopting ME hypotheses. The
	second one is a class that, for example,
	could be interpreted in
	terms of mixed polarities in a single pixel of \textit{HINODE},
	i.e., with a two magnetic component model atmosphere.
	Recently, \citet{AseR10spw6} put forward
	an inversion procedure based on Bayesian techniques.
	This procedure follows directly from the approach
	adopted in \citet{AseR09}, based exclusively on ME hypothesis.
	The new code is able to automatically
	recognize the simplest model, from those
	provided by the user that can reproduce the examined polarization
	profiles. In other words,
	the code recognizes the minimum degree of complexity
	of a model atmosphere needed to reproduce the
	observations.
	This means that
	different inversion hypotheses can be used in a
	single inversion procedure.
	From our point
	of view such an approach is extremely promising because
	it allows to avoid an a-priori choice of hypotheses
	for the description of the solar photosphere
	that, in some cases, can turn out to be wrong in reproducing
	observations.	
	
	Among the profile shapes in Fig.~\ref{fig1} many single lobe
	profiles can be recognized. Namely, according to the definition
	adopted in \citet{Ste00}, $17.6$~\% of the analyzed profiles are
	blue-lobe-only, while $9.6$~\% are red-lobe-only.
	Such extremely asymmetric profiles have been
	the object of recent studies
	\citep[][]{GroD88,GroD89,SanA96,GroD00,Ste00}.
	In \citet{GroD00}, the authors pointed out
	two models to produce such profiles:
	the canopy model and the embedded magnetic flux tube model.
	Both models are	characterized by
	a discontinuity along the LOS between the field-free component
	and the magnetized one.
	The main conclusion of this work is that 
	both models are able to reproduce the observed asymmetries.
	More in detail, the canopy model needs an appropriate
	temperature stratification to produce strong asymmetries
	while the embedded magnetic flux tube model is able to
	produce them with a standard quiet
	Sun temperature stratification.
	
	The importance of the thermal properties of
	the solar atmosphere in the formation of strongly
	asymmetric profiles is also highlighted in
	\citet{Ste00} as one of the
	main conclusions of the paper
	in which a canopy configuration
	under ME hypotheses is considered.
	From this study it follows that
	a temperature inversion, causing an
	emission in the magnetized layer,
	is needed at the magnetopause
	of the canopy to produce blue-lobe-only profiles
	in upflow regions as observed in \citet{Sig99}
	and in our analysis. Besides this
	case, also $Q$-like profiles can
	be produced thanks to emission processes
	in the magnetized layer.
	In \citet{Ste00} the author reported
	a complete atlas of asymmetric profiles
	produced with a canopy configuration.
			
	In \citet{Ste00} Stokes~$V$ profiles emerging
	from a micro-loop smaller than the resolution
	element of the observations are also discussed.
	With such a figure one can explain the formation of
	both blue-lobe-only and red-lobe-only profiles and the
	correlation of these two families with the photospheric
	dynamics.
	The micro-loop model implicates the presence of a loop
	in the resolution element, i.e., two opposite polarities
	corresponding to the legs (footpoints) of the loop.
	Considering this figure, $Q$-like profiles can be
	easily produced, moreover, it can be shown that a
	suppression of the blue (red) lobe in the emerging profiles
	in correspondence of downflows (upflows) can be
	obtained. Similarly to what was found
	for the canopy model, one can imagine the definition
	of an atlas of asymmetric profiles from the micro-loop model.
	Any modification of the internal dynamics of the micro
	loop can influence the lobe cancellation leading either to
	a $Q$-like profile or to a profile partially dominated
	by a blue/red lobe as those we found in the
	classification.
	
	From our discussion it follows 
	that adopting either a model with 
	a magnetopause along the LOS or a micro-loop model
	one could, in principle, reproduce
	a large part of the profiles observed
	in the quiet Sun by SOT/SP.
	
	We have to point out that
	the micro-loop model was introduced by \citet{Ste00}
	for the interpretation of $1$~arcsec angular resolution
	spectropolarimetric data presented in \citet{Sig99}. This is an important point because
	the micro-loop is supposed to be contained in the resolution
	element of the observation. SOT/SP angular resolution is $0.3$~arcsec,
	i.e., three times smaller than that of the dataset
	analyzed in \citet{Sig99},
	but still many strong asymmetries compatible
	with the micro-loop model emerge in quiet
	Sun polarimetric measurements.	
	The possible presence of micro-loops in the
	\textit{HINODE} resolution element is of big interest
	for the solar community. Indeed, such a figure
	is in close connection with many recent works that have
	been dedicated to the study of loop
	emergence events in the quiet Sun as detected by
	\textit{HINODE} \citep[e.g., ][]{Cen07,OroS08,MarG09,MarG10}.
	As an example, in \citet{OroS08} Stokes~$V$ profiles observed in
	correspondence with a loop emergence event are reported;
	at the beginning of the process a single blue-lobe
	profile is found.
	In the same work the authors presented preliminary
	inversions of single lobe profiles performed through
	a model atmosphere characterized by a strong discontinuity
	in the stratification \citep[as in a canopy model of ][]{GroD00}.
	This analysis and that by \citet{Lou09}
	are based on SIRJUMP, a modified version of SIRGAUSS,
	which is able to deal with discontinuities along the LOS.
	In-depth analyses of single-lobe
	profiles observed by \textit{HINODE} via the SIRJUMP
	code are ongoing \citep[][]{SaiD10spw6}.
	
	A last possible scenario for reproducing the variety
	of Stokes~$V$ asymmetries observed in the quiet Sun is the
	MISMA scenario, i.e., a MIcro-Structured
	Magnetized Atmosphere \citep{SanA96}. In this model
	the solar atmosphere is thought to be composed
	by many optically thin magnetized components with different
	plasma dynamics embedded in a field-free gas.
	These components naturally arise from the interplay between convection
	and magnetic fields \citep[e.g., ][]{Cat99,PieG09}.
	\citet{GroD00} correctly considered MISMAs as
	a generalization of the model they proposed
	when considering many transients
	along the LOS. Here we point out
	that in the MISMA formulation also mixed
	polarity (loop-like) models are required \citep{SanALit00}.
	Namely, they are needed to explain the $Q$-like
	profiles and many of the one-lobe profiles.
	The MISMA hypothesis has been recently adopted to
	provide the first interpretation of the
	asymmetries of \textit{HINODE} quiet Sun profiles \citep{Vit11}.
	One of the main results presented in this work
	is the plenitude of mixed polarity pixels, i.e., $25$~\% of
	the IN profiles. This value is three times
	the abundance of $Q$-like profiles found here.
	However, mixed polarity profiles could be present in principle in many
	other classes that present extreme asymmetries.
	As an example, \texttt{class~1}
	($5.2$~\%) could contain mixed polarity pixels.
	
	From the above discussion it follows
	that different models/hypotheses can be used
	for the interpretation of the strongly asymmetric
	polarization profiles observed by SOT/SP and, usually,
	different models are associated to different
	inversion codes. Among these, the MISMA
	inversion code \citep[][]{SanA97} is the only one so far used
	for the interpretation of asymmetries in SOT/SP data, and it turns out to be
	suitable for the interpretation of all the profile shapes highlighted
	in Fig.~\ref{fig1} \citep[][]{Vit11}. From our point of view,
	different and independent interpretations
	from different analysis methods can be
	a key factor to clarify the physical picture
	behind the asymmetries recently measured in the
	quiet Sun. The approach adopted by
	\citet{AseR10spw6} clearly goes in this
	direction.
	
	By comparing the results of our classification
	with previous studies on the shapes of Stokes $V$
	profiles in the quiet Sun we note that the amount
	of strongly asymmetric profiles seems to increase
	with the spatial resolution of observations.
	Here we refer to \cite{Sig99} and \cite{SanALit00}
	for a comparison. In both papers,
	spectropolarimetric observations were performed through
	the Advanced Stokes Polarimeter (ASP) with a
	spectral sampling of approximately $1.2$~pm pixel$^{-1}$,
	a polarimetric sensitivity on the order of $10^{-4}$, and
	an angular resolution of about $1$~arcsec. From their
	respective analyses the authors obtained similar results
	on the abundance of strongly asymmetric profiles.
	Referring to \citet[][Fig.~4]{SanALit00} one concludes
	that roughly $17$~\% of the analyzed profiles
	considerably deviate from antisymmetric shapes; in
	\citet{Sig99} the percentage is of about $10$~\% in the
	quiet Sun. These abundances are at least a half the
	abundance we obtained \citep[around $35$~\%, in
	agreement with][who analyzed the same dataset]{Vit11}.
	We note that both the spectral sampling and the polarimetric
	sensitivity provided by ASP are suitable to
	go into the details of Stokes $V$ shapes. Moreover,
	the methods adopted by the authors to study the Stokes $V$
	shapes can be regarded as reliable. In spite
	of this, from SOT/SP measurements with $0.3$~arcsec angular
	resolution and $10^{-3}$ polarimetric sensitivity
	we find an increase in the abundance of strongly
	asymmetric profiles. This result is not surprising
	if we refer to the typical dimension of the patches
	associated to the IN families of our classification,
	i.e., few tenths of $1$~arcsec. From this it
	follows that strong asymmetries are usually found
	over small scales. This fully agrees with
	the conclusions of \citet{Sig99}, in which the authors
	report that \textit{``The broad scatter in the $V$-parameters
	for the weakest $V$ signals can be interpreted as a dramatic
	increase in the dynamic behaviour with decreasing size
	(and fill factor) of magnetic elements''}.
	It is very important to remember that
	most of the IN pixels were overlooked in our analysis
	because of the selection criteria on the Stokes $V$
	amplitude. This could in principle strongly affect
	the dimension of the patches associated to each IN
	family. Indeed, by considering Stokes signals
	close to the SOT/SP noise level would considerably increase
	the number of asymmetric profiles and affect the
	properties of IN family patches.
	
	Another result in agreement with \citet{Sig99}
	is the difference in the dynamical properties of the
	blue-lobe family and red-lobe family, respectively.
	With respect to this, one has to remember that the dynamical
	properties in Fig.~\ref{fig4} were derived exclusively
	from Stokes $I$ profiles
	so that it is probably too optimistic to state that
	different dynamical properties of the non magnetized plasma can
	give rise to different Stokes $V$ shapes. Rather, we can
	think about different field-free dynamics associated to
	certain atmospheric configurations that are typical either of
	granules or intergranular lanes (as discussed above).
			
	As a final remark, we now discuss two possible applications
	of the set of \texttt{k-means} profiles reported in Fig.~\ref{fig1}.
	The first one is the definition of a set of profiles
	that can be used to check whether the profiles synthesized
	from MHD simulations can
	be confidently considered to be representative of the profiles
	observed by \textit{HINODE}. As an example, in \cite{VogSch07}
	MURaM local dynamo simulation magnetic fields turn out to be
	extremely variable in polarity over very
	small scales. This implies the presence of mixed
	polarities over scales comparable to 
	\textit{HINODE} resolution, possibly producing  asymmetric profiles 
	like those in SOT/SP observations.
	This is an important and easy check that has not
	been considered so far when analyzing Stokes profiles
	produced from MURaM simulations. 
	Such a check is very important because many
	MURaM snapshots have been used to test inversion
	codes employed for extensive analyses of
	\textit{HINODE} data. As an example, in
	\citet{OroS07b} the authors considered three
	MURaM snapshots to test their ME
	inversion code \citep{OroSDelT07}. The three snapshots
	were selected to be unipolar and with
	different values of the unsigned flux, namely,
	$10$~G, $50$~G, and $200$~G, so to take into account
	different magnetic regimes observable in the
	quiet Sun. To reproduce \textit{HINODE} observations
	the finite spatial and spectral resolutions of
	SOT/SP were considered through a degradation of the synthesized
	data. The $10^{-3}$ polarimetric sensitivity
	was also taken into account to define the noise
	level of the synthesized observations.
	The strategy adopted by the authors can be
	regarded to be complete and rigorous, but at
	the same time a question is still open.
	Are the profiles produced by the snapshots
	selected by \citet{OroS07b} an exhaustive
	set of profiles for the description of
	\textit{HINODE} measurements? 
	One could answer the question by comparing the synthetic
        profiles used by \citet{OroS07b} with those portrayed
	in Fig.~\ref{fig1}.
        If they agree, then it would have two important consequences.
        First, the ME code has been tested
	on a complete set of profiles which is
	representative of real observations so, as a
	direct consequence, the test guarantees that
	an analysis of SOT/SP data performed through
	a ME code is reliable. The second one is
	that MURaM snapshots are able to reproduce
	the physics producing \textit{HINODE} SOT/SP
	measurements.
	
\section{Conclusion}
\label{Conc}
	We present the results of the \texttt{k-means}
	classification of \textit{HINODE} SOT/SP Stokes~$V$
	profiles of \ion{Fe}{i}~$630.15$~nm and \ion{Fe}{i}~$630.25$~nm
	observed in the quiet Sun. The classification
	procedure is automatic and unsupervised, and it
	is able to highlight the typical circular
	polarization measurements retrieved by SOT/SP,
	and to organize them in classes.
	Here we analyze the profiles from a $302\times162$~arcsec$^2$
	portion of quiet photosphere. This dataset,
	consisting of $535465$ profiles, can be regarded
	as a complete sample of
	polarization measurements performed
	at $0.3$~arcsec angular resolution and $10^{-3}$
	polarimetric sensitivity in the quiet Sun.
	
	A large variety of profile shapes
	emerges from the quiet Sun. These
	shapes can be typically classified in $35$
	\texttt{k-means} classes (Fig.~\ref{fig1}).
	Referring to the major features of the
	\texttt{k-means} classes, we were able to
	organize them in
	six families, namely, network, blue-lobe,
	red-lobe, $Q$-like, asymmetric, and
	antisymmetric.
	
	Network and IN profile classes are well
	separated in shape. Network profiles represent
	approximately $28$~\% of the analyzed profiles.
	
	Asymmetric profiles are very
	common in the quiet Sun, namely,
	about $93$~\% of the analyzed
	quiet Sun profiles.
	Strongly asymmetric profiles are found
	in the IN. They can be organized in
	three families: \textit{i)}
	blue-lobe, \textit{ii)} red-lobe, and
	\textit{iii)} $Q$-like profiles
	\citep[similarly to the definition in][]{Ste00}.
	These represent about $34$~\% of the analyzed profiles.
	Each family of profiles is associated to a certain photospheric
	dynamics as inferred from the line-core shift
	of the intensity profiles.
	Blue-lobe profiles are in upflow regions,
	red-lobe ones are in downflows, and $Q$-like
	profiles are in downflows; these
	dynamical properties are probably related
	to certain magnetic configurations preferentially
	associated with granules or with intergranular lanes.

	The classes associated to
	the network are found to be spatially coherent
	over tens of arcsec$^2$. IN classes
	are found to be coherent over small scales
	(i.e., $<1$~acrsec$^2$). From this we conclude
	that the processes producing strong asymmetries
	in Stokes $V$ profiles take place over
	small scales, comparable to SOT/SP angular
	resolution.
	
	We put forward a possible application
 	of the set of \texttt{k-means} cluster profiles
 	(Fig.~\ref{fig1}). They contain all the Stokes~$V$ 
	shapes observed in the quiet Sun at $0.3$~arcsec
	resolution and with maximum absolute amplitude
	larger than $4.5\times10^{-3}$, therefore, they can be used as a standard
 	reference to test inversion codes and to check
 	the validity and completeness of modern MHD simulations.
 	
 	From the abundance of
 	asymmetric profiles we conclude that
 	ME inversions cannot be considered to be exhaustive
 	for the description of the properties of
 	the quiet Sun photosphere.
 	Indeed, Stokes~$V$ profiles observed by \textit{HINODE}
 	SOT/SP in the quiet Sun contain unique
	information about the stratification of the
	solar photosphere that so far have been overlooked
	in ME analyses. The work of \citet{Vit11}
	is the first attempt to try to extract this information. 
 	However, as shown in \S~\ref{Disc}, several alternative
 	interpretations
 	are possible through hypotheses on the underlying
	atmospheres. These alternatives should be
	sought to outline the spectrum of physical scenarios
	for the description of the quiet Sun
 	polarization measurements.
	Those features common to all inversions will be
	regarded as robust results, whereas inconsistencies
	must be studied and cleared out. 
	
	\acknowledgements
	The authors acknowledge the referee Luis Bellot Rubio for his
	useful comments and suggestions, which helped to improve the
	manuscript.
	\textit{HINODE} is a Japanese mission developed and launched by ISAS/JAXA,
	collaborating with NAOJ as a domestic partner, NASA and STFC (UK) as
	international partners. Scientific operation of the \textit{HINODE} mission is
	conducted by the \textit{HINODE} science team organized at ISAS/JAXA. This team
	mainly consists of scientists from institutes in the partner countries.
	Support for the post-launch operation is provided by JAXA and NAOJ (Japan),
	STFC (U.K.), NASA, ESA, and NSC (Norway).
	Fruitful discussions with D.~M{\"u}ller and N.~Vitas are acknowledged.
	JSA acknowledges the support provided by the Spanish Ministry 
	of Science and Technology through project AYA2007-66502, as well as 
	by the EC SOLAIRE Network (MTRN-CT-2006-035484). This work
	was partially supported by ASI grant n.I/015/07/0ESS.
	



\end{document}